\documentclass[aip,jap,amsmath,amssymb,preprint,
]{revtex4-1}
\usepackage{graphicx}
\usepackage{bm}
\usepackage{subfigure}
\usepackage[T1]{fontenc}
\usepackage[utf8]{inputenc}
\usepackage{lmodern}
\newcommand{\be}{\begin{equation}}
\newcommand{\ee}{\end{equation}}
\newcommand{\bea}{\begin{eqnarray}}
\newcommand{\eea}{\end{eqnarray}}

\begin{document}
\title{Exact Intrinsic Localized Excitation of an Anisotropic Ferromagnetic 
Spin Chain in External Magnetic Field with Gilbert Damping, Spin Current and 
$\mathcal{PT}$-Symmetry}
\author{M. Lakshmanan}%
 \email{lakshman.cnld@gmail.com}
\affiliation{Centre for Nonlinear Dynamics, School of Physics, Bharathidasan University, Tiruchirappalli - 620 024, India}%

\author{Avadh Saxena}%
\email{avadh@lanl.gov}
\affiliation{Theoretical Division and Center for Nonlinear Studies, Los
Alamos National Laboratory, Los Alamos, NM 87545, USA}%
\begin{abstract}
We obtain the exact one-spin intrinsic localized excitation in an anisotropic Heisenberg 
ferromagnetic spin chain in a constant/variable external magnetic field with Gilbert 
damping included.  We also point out how an appropriate magnitude spin current 
term in a spin transfer nano-oscillator (STNO) can stabilize the tendency towards 
damping. Further, we show how this excitation can be sustained in a recently suggested 
$\mathcal{PT}$-symmetric magnetic nanostructure.  We also briefly consider more general 
spin excitations. 
\end{abstract}
\maketitle
\section{Introduction}
The study of dynamics of classical Heisenberg ferromagnetic spin chain with anisotropic 
interaction is of considerable importance in applied magnetism \cite{Hiller, ML1} and 
from application point of view \cite{Fert, Yang}.  While several continuum versions are 
known to be completely integrable soliton systems \cite{ML2, Naka, Sklyanin}, such as 
the isotropic case, no discrete integrable case is known in the literature, except for a modified 
version, namely the Ishimori lattice \cite{Ishi}.  On the other hand, the present authors 
\cite{ML3} have shown the existence of several classes of exact solutions in terms of 
Jacobian elliptic functions which exist for the case of the discrete lattice including onsite 
anisotropy and external magnetic field.  Identifying such interesting classes of solutions 
and their relevance in the context of appropriate physical situations constitute one of 
the important areas of investigation in spin dynamics \cite{ML3, Zabel}. 

From another point of view, occurrence of intrinsic localized breathers/oscillations in 
suitable anisotropic ferromagnetic spin chains is of practical relevance \cite{Sievers, Zolo} 
and is being explored for the past several years.  Apart from many numerical studies, in 
recent times the present authors and Subash \cite{Subash1} have obtained explicit 
analytical solutions for the Heisenberg anisotropic spin chain with additional onsite 
anisotropy and constant external magnetic field corresponding to excitations of one, 
two and three spins and also investigated their stability.  Additionally, relevant situations 
were pointed out where such excitations can be physically identified. 

In recent times, one has also seen that at nanoscale level spin transfer nano-oscillator 
(STNO) \cite{Bert, Georges}, which essentially consists of a trilayer structure of two 
nanoscale ferromagnetic films separated by a non-ferromagnetic but conducting layer,  
can lead to switching of spin angular momentum directions and allow for the generation 
of microwave oscillations \cite{Subash2, Turtle}. The ferromagnetic film even when it 
is homogeneous is dominated by anisotropic interactions besides the presence of 
external magnetic fields (both dc and ac) and spin current terms. The equation of motion 
defining the evolution of the spins 
is the Landau-Lifshitz-Gilbert-Slonczewski (LLGS) equation \cite{LLGS} where the spin 
current term is given by the Slonczewski form.  One notices that the LLGS equation is a 
simple generalization of the Landau-Lifshitz-Gilbert (LLG) equation which describes the 
nonlinear magnetization dynamics in bulk materials as in the case of ferromagnetic lattices.  
Then it becomes important to ask what is the influence of spin current term on the spin 
excitations, particularly intrinsic localized oscillations (ILOs) and identify the conditions 
under which damping effect can be off-set by the spin current term. 

From yet another point of view, one may consider the possibility of designing a 
$\mathcal{PT}$-symmetric ferromagnetic nanoscale device by considering two nano-film 
structures interspersed by a nonmagnetic but conducting thinner layer (i.e. a sandwich 
structure) as suggested by Lee, Kottos and Shapiro \cite{Kottos} very recently.  These 
authors have proposed a class of synthetic magnetic nanostructures which utilize natural 
dissipation (loss) mechanisms along with suitable chosen gain mechanism so as to control 
the magnetization dynamics. We will also explore how the spin ILOs can be identified in 
these structures.  

In this paper we deduce an explicit one-spin excitation in an anisotropic ferromagnetic 
lattice (without onsite anisotropy, to start with) in the presence of external magnetic field 
and explore the effect of spin current term to maintain the oscillatory nature of the spin 
excitation.  We then point out how this can be generalized to more general spin 
excitations and in $\mathcal{PT}$-symmetric nanostructures. 

The organization of the paper is as follows.  In Sec. 2 we deduce the dynamical equation for an 
anisotropic ferromagnetic spin in the presence of external magnetic field and set up the 
appropriate equation for a one-spin excitation in the presence of Gilbert damping.  In 
Sec. 3, we deduce the explicit one-spin excitation including the damping effect and 
analyze how the spin excitation gets affected by the damping.  In Sec. 4, we incorporate 
the spin current term and point out how an appropriate strength of spin current can 
off-set the effect of damping so as to control the spin oscillations.  In Sec. 5, we point out 
how the above analysis can be extended to a $\mathcal{PT}$-symmetric nanostructure.  
We briefly indicate how this study can be extended to consider more general spin 
excitations in Sec. 6.  Finally in Sec. 7, we present our conclusions.
\section{Dynamics of the anisotropic spin chain and one-spin excitation}

Considering the evolution of spins of a one-dimensional anisotropic Heisenberg ferromagnetic 
spin chain modeled by the Hamiltonian \cite{Zolo} 
\be 
H=-\sum^N_{\{n\}} (AS_n^xS_{n+1}^x + BS_n^yS_{n+1}^y + CS_n^zS_{n+1}^z) 
-D\sum_n(S_n^z)^2 - \vec{\mathcal{H}}\cdot\sum_n\vec{S}_n \, ,
\ee
where the spin components $\vec{S}_n = (S_n^x, S_n^y, S_n^z)$ are classical  unit vectors 
satisfying the constant length condition 
\be 
(S_n^x)^2 + (S_n^y)^2 + (S_n^z)^2 = 1,  ~~~ n=1,2, ..., N \, . 
\ee
Here $A$, $B$, and $C$ are the exchange anisotropy parameters, $D$ is the onsite 
anisotropy parameter and the external magnetic field $\vec{\mathcal{H}}=(\mathcal{H},0,0)$ is chosen along 
the x-axis for convenience.  By introducing the appropriate spin-Poisson brackets and 
deducing the LLG spin evolution equation one can obtain the equation for the spin 
lattice (1) as 
\be 
\frac{d\vec{S}_n }{dt} = \vec{S}_n\times\vec{\mathcal{H}}_{eff} + \alpha\vec{S}_n\times(\vec{S}_n\times\vec{\mathcal{H}}_{eff}) \, ,
\ee 
where 
\be 
\vec{\mathcal{H}}_{eff} = A (S^x_{n+1}+S^x_{n-1})\hat{i} + B (S^y_{n+1}+S^y_{n-1})\hat{j} + C(S^z_{n+1}+S^z_{n-1})\hat{k} + 2D S_n^z\hat{k} + \vec{\mathcal{H}} \, , 
\ee 
and $\alpha$ is the Gilbert damping parameter.  In component form Eq. (3) with Eq. (4) 
reads as  
\bea 
&&\frac{d{S}^x_n }{dt} = CS_n^y(S^z_{n+1}+S^z_{n-1}) -BS_n^z(S^y_{n+1}+S^y_{n-1}) -2D S_n^yS_n^z 
+\alpha\Big[BS_n^yS_n^z(S_{n+1}^y+S_{n-1}^y)  \nonumber \\ &&
- A(S_{n+1}^x+S_{n-1}^x)((S_n^y)^2+(S_n^z)^2)  
+CS_n^zS_n^x(S_{n+1}^z+S_{n-1}^z) - 2DS_n^x(S_n^z)^2 -\mathcal{H}((S_n^x)^2+(S_n^z)^2)\Big] \, ,  
\eea  
\bea 
&&\frac{d{S}^y_n }{dt} = AS_n^z(S^x_{n+1}+S^x_{n-1}) -CS_n^x(S^z_{n+1}+S^z_{n-1}) + 2D S_n^xS_n^z + \mathcal{H} S_n^z +\alpha\Big[CS_n^zS_n^y(S_{n+1}^z+S_{n-1}^z)  \nonumber \\ &&
- B((S_n^x)^2+(S_{n}^z)^2)(S_{n+1}^y+S_{n-1}^y)  
+AS_n^xS_n^y(S_{n+1}^x+S_{n-1}^x) - 2DS_n^y(S_n^z)^2 + \mathcal{H} S_n^xS_n^y\Big] \, , 
\eea 
\bea 
&&\frac{d{S}^z_n }{dt} = BS_n^x(S^y_{n+1}+S^y_{n-1}) -AS_n^y(S^x_{n+1}+S^x_{n-1}) - \mathcal{H} S_n^y 
+\alpha\Big[AS_n^xS_n^y(S_{n+1}^x+S_{n-1}^x)  \nonumber \\ &&
- C((S_n^x)^2+(S_n^y)^2)(S_{n+1}^z+S_{n-1}^z)  
+BS_n^yS_n^z(S_{n+1}^y+S_{n-1}^y) + 2DS_n^z((S_n^x)^2+(S_n^y)^2) + \mathcal{H} S_n^xS_n^z\Big] \, . 
\eea 
Now looking for the one spin excitation for (1) as
\be 
\vec{S}_n = ... , (1,0,0), (1,0,0), (S^x_i(t), S^y_i(t), S^z_i(t)), (1,0,0), (1,0,0), ... \, , 
\ee 
where we have used $n$ to denote a general spin in the lattice and used $i$ to specify the 
localized spin excitation, and redesignating  $(S_i^x(t) , S_i^y(t), S_i^z(t))$ as $(S_0^x(t) , S_0^y(t), S_0^z(t))$, the equation of motion (LLG 
equation) for the excited spin can be given as 
\be 
\frac{d{S}^x_0 }{dt} = -2DS_0^yS_0^z -\alpha\Big[(2A+\mathcal{H})((S_0^y)^2+(S_0^z)^2) + 2DS_0^x(S_0^z)^2\Big] \, , 
\ee 
\be 
\frac{d{S}^y_0 }{dt} = (2A+\mathcal{H})S_0^z + 2DS_0^xS_0^z + \alpha\Big[(2A+\mathcal{H})S_0^xS_0^y - 2DS_0^y(S_0^z)^2 \Big] \, , 
\ee 
\be 
\frac{d{S}^z_0 }{dt} = -(2A+\mathcal{H})S_0^y  + \alpha\Big[(2A+\mathcal{H})S_0^xS_0^z + 2D((S_0^x)^2 + (S_0^y)^2)S_0^z\Big] \, .  
\ee 
Note that from Eqs. (9) - (11), one can check that  
\be 
S_0^x \frac{d{S}^x_0 }{dt} +S_0^y \frac{d{S}^y_0 }{dt} + S_0^z \frac{d{S}^z_0 }{dt} = 0 \, , 
\ee 
so that $\vec{S}^2 = (S_o^x)^2 + (S_0^y)^2 + (S_0^z)^2 = Constant = 1$ is conserved. 

Next, further confining to the case where the onsite anisotropy vanishes, $D=0$, we have 
the LLG equation for the one-spin excitation,
\be 
\frac{d{S}^x_0 }{dt} =  -\alpha(2A+\mathcal{H})(1-(S_0^x)^2) \, , 
\ee 
\be 
\frac{d{S}^y_0 }{dt} = (2A+\mathcal{H})S_0^z  + \alpha(2A+\mathcal{H})S_0^xS_0^y \, , 
\ee 
\be 
\frac{d{S}^z_0 }{dt} = -(2A+\mathcal{H})S_0^y  + \alpha(2A+\mathcal{H})S_0^xS_0^z \, ,
\ee 
with the constraint $\vec{S}^2 = (S_o^x)^2 + (S_0^y)^2 + (S_0^z)^2 = 1$ . 
The system (13) - (15) can be exactly solved as shown below. 

\section{Explicit one-spin excitation} 

Now the above system of nonlinear differential equations can be straightforwardly solved. 
Integrating (14) we obtain 
\be 
S_0^x(t) = \frac{c^2 e^{-2\alpha(2A+\mathcal{H})t} -1}{c^2 e^{-2\alpha(2A+\mathcal{H})t}+1} \, , 
\ee 
where $c$ is an arbitrary constant.  We also note that when $\alpha=0$, that is no damping, 
$S_0^x (t) = (c^2-1)/(c^2+1) = const = \sqrt{1-a^2}$ as noted in ref. [13], Eq. (11).  Also we 
note that $S_0^x (0) = (c^2-1)/(c^2+1)$ and $S_0^x (\infty)=-1$, indicating a switching from 
a given initial value to the other ground state, $S_0^x=-1$.  

To find $S_0^y$ and $S_0^z$, we proceed as follows.  Considering Eq. (14) and differentiating 
once with respect to $t$ on both sides to obtain $(d^2S_0^y/dt^2)$, after making use of the forms 
of $(dS_0^x/dt)$ and $(dS_0^z/dt)$ from (13) and (15), respectively, we have 
\be 
\frac{d^2S_0^y}{dt^2} = -(2A+\mathcal{H})^2(1+\alpha^2)S_0^y + 2\alpha(2A+\mathcal{H})^2S_0^xS_0^z 
+ 2\alpha^2(2A+\mathcal{H})^2(S_0^x)^2S_0^y \, , 
\ee 
so that 
\be 
S_0^z(t) = \frac{1}{2\alpha(2A+\mathcal{H})^2S_0^x} \left[\frac{d^2S_0^y}{dt^2}+(2A+\mathcal{H})^2(1+\alpha^2)S_0^y -2\alpha^2(2A+\mathcal{H})^2(S_0^x)^2S_0^y  \right] \, . 
\ee 
Also from (14) we can write 
\be 
S_0^z(t) = \frac{1}{2A+\mathcal{H}}\left[\frac{dS_0^y}{dt} -\alpha S_0^xS_0^y   \right] \, .
\ee
Equating the right hand sides of (18) and (19), we obtain 
\be 
\frac{d^2S_0^y}{dt^2} = -2\alpha(2A+\mathcal{H})S_0^x \frac{dS_0^y}{dt} + (2A+\mathcal{H})^2(1+\alpha^2)S_0^y = 0 \, . 
\ee 
After a standard transformation and two integrations (as indicated in Appendix A), we can explicitly write the solution for $S_0^y$ as
\be 
S_0^y = \frac{c\exp(-\alpha(2A+\mathcal{H})t)} {c^2 \exp({-2\alpha(2A+\mathcal{H})t})+1}~ \hat{a}\cos(\Omega t + \delta) \,  , 
~~~ \Omega = 2A+\mathcal{H} \, , 
\ee
where $\hat{a}$ is an arbitrary constant.  
Also from (14) we have 
\be 
S_0^z(t) = \frac{1}{(2A+\mathcal{H})}\left[\frac{dS_0^y}{dt}-\alpha S_0^xS_0^y  \right] = -\hat{a}\sin(\Omega t+\delta) \frac{c 
e^{-\alpha(2A+\mathcal{H})t}}{c^2e^{(-2\alpha(2A+\mathcal{H})t} + 1} \, . 
\ee 
Now in order to fix the constant $\hat{a}$ we demand that the spin length constraint $(S_0^x)^2 + 
(S_0^y)^2 + (S_0^z)^2 =1$ be valid.  This leads to 
\be 
\hat{a}^2=4 ~~or~~ \hat{a}=2 \, , 
\ee 
so that we have now the complete solution of the excited spin as 
\be 
S_0^x(t) = \frac{c^2 e^{-2\alpha(2A+\mathcal{H})t} -1}{c^2 e^{ -2\alpha(2A+\mathcal{H})t} + 1}  \, , 
\ee 
\be 
S_0^y(t) = \frac{2c e^{-\alpha(2A+\mathcal{H})t}}{c^2 e^{ -2\alpha(2A+\mathcal{H})t} + 1} \cos[(2A+\mathcal{H})t+\delta]  \, , 
\ee 
\be 
S_0^z(t) = -\frac{2c e^{-2\alpha(2A+\mathcal{H})t} }{c^2 e^{ -2\alpha(2A+\mathcal{H})t} + 1} \sin[(2A+\mathcal{H})t+\delta] \, . 
\ee 
Note that the arbitrary constant corresponding to the undamped case ($\alpha=0$) is
\be 
\hat{a} = \frac{c^2-1}{c^2+1} \, . 
\ee
It is obvious from the above that for $\alpha=0$, $S^{x}_{0}=constant=\frac{c^{2}-1}{c^{2}+1}$, while $S^{y}_{0}$ and $S^{z}_{0}$ are periodic functions of $t$. In this case Eqs. (14)-(15) are linear in $S^{y}_{0}$ and $S^{z}_{0}$ so that the perturbation around the origin in the ($S^{y}_{0}-S^{z}_{0}$) plane admits pure imaginary eigenvalues. When $\alpha\ne 0$, they get damped as shown in Fig. 1 corresponding to the explicit forms (24)-(26). Note that in the above we have assumed the external magnetic field to be a constant in time. 
However, even in the case where the field is a variable function of time, say 
\be 
\mathcal{H}(t) = h_0 + h_1\cos\omega t \, , 
\ee 
where $h_0$, $h_1$ and $\omega$ are constants, we observe from the equations of motion of the 
spin components (13) - (15), that $\mathcal{H}$ occurs always as a linear combination $2A+\mathcal{H}(t) = 
(2A+h_0+h_1\cos\omega t)$.  Therefore by redefining the time $(2A+\mathcal{H})t$ as 
\be 
\tau = (2A + h_0)t - h_1\omega\sin\omega t \, , 
\ee 
all the previous analysis goes through.  The final spin excitations are of the same form as 
(24) - (26) but with the transformed time variable given by Eq. (29). 
\begin{figure}[htb!]
\centering
\subfigure{\includegraphics[width=0.32\linewidth]{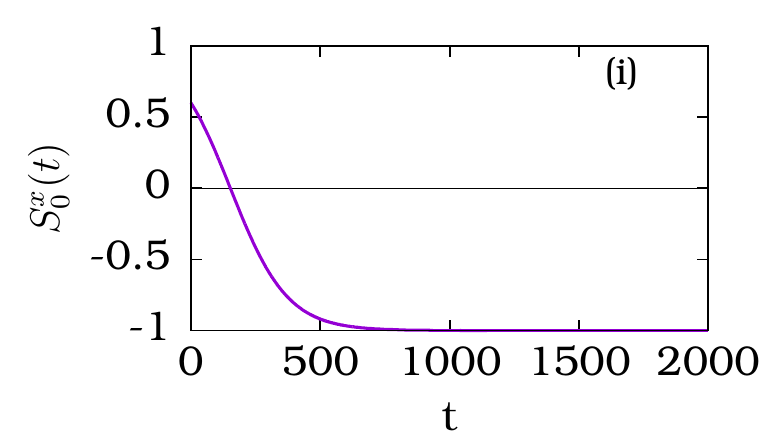}}
\subfigure{\includegraphics[width=0.32\linewidth]{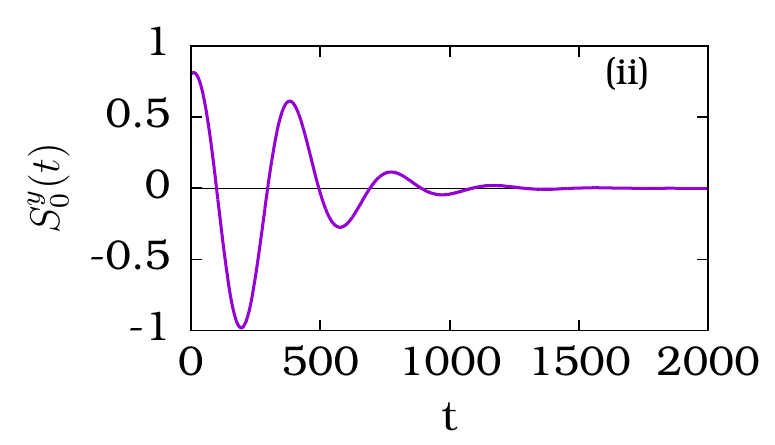}}
\subfigure{\includegraphics[width=0.32\linewidth]{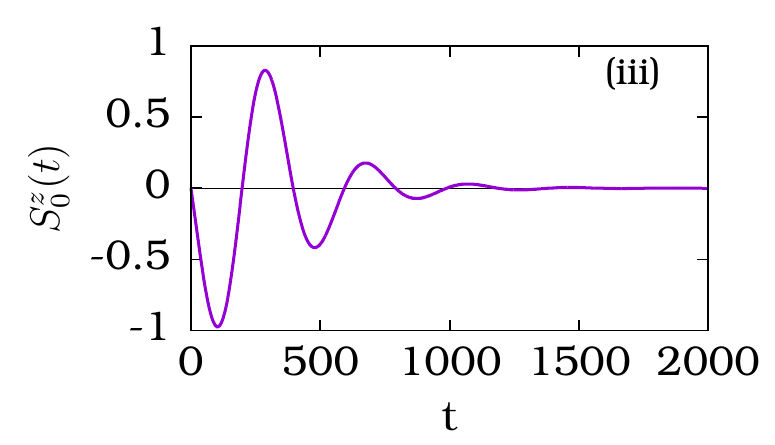}}
\caption{Damped spin excitation: One-spin excitation (Eqs. (24)-(26)) showing the three spin components for the damped cases ($\alpha=0.005$).} 
\label{fig:numplot}
\end{figure}
\section{Effect of Slonczewski spin current} 
Next we consider the influence of spin current term in a trilayer structured STNO (see Fig. 2), 
where we consider the excitation of a single spin of magnetization in the outer uniformly 
magnetized layer under anisotropic interaction and external magnetic field in the presence 
of spin current.  The corresponding spin excitation is given by the Landau-Lifshitz-Gilbert-Slonczewski 
equation for the spin as 
\begin{figure}[htp!]
  \begin{center}
    \includegraphics[width=0.3\linewidth]{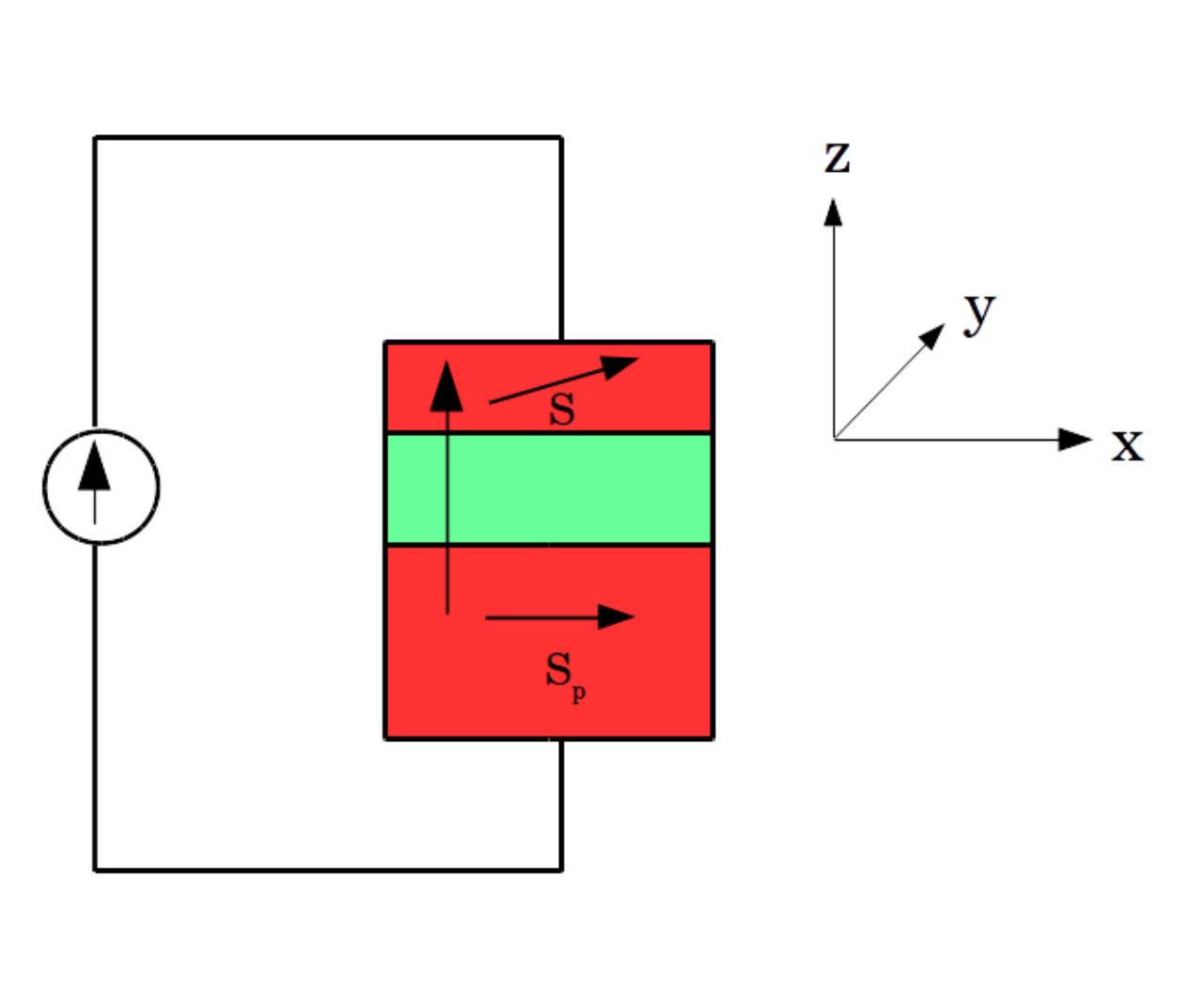}
    \caption{A schematic representation of STNO.}
    \label{fig:stno}
  \end{center}
\end{figure}
\be 
\frac{d\vec{S}_n}{dt} = \vec{S}_n\times \vec{\mathcal{H}}_{eff} + \alpha \vec{S}_n\times(\vec{S}_n\times \vec{\mathcal{H}}_{eff})  
+j \vec{S}_n\times(\vec{S}_n\times\vec{S}_p) \, , 
\ee 
where $\vec{\mathcal{H}}_{eff}$ is the effective field given by Eq. (4) and $j$ is the magnitude of the 
spin current and the polarization vector $\vec{S}_p$ is 
\be 
\vec{S}_p = (1,0,0) \, , 
\ee
corresponding to the flow of electrons in the x-direction.  Consequently, 
\be 
\vec{S}_n\times(\vec{S}_n\times\vec{S}_p) = \vec{S}_n(\vec{S}_n \cdot \vec{S}_p)-\vec{S}_p =- ((S_n^y)^2+(S_n^z)^2 )\hat{i} + S_n^xS_n^y\hat{j} 
+ S_n^xS_n^z\hat{k} \, , 
\ee 
where $(\hat{i}, \hat{j}, \hat{k})$ form the unit orthonormal trihedral.  As a result, the equations 
for the one-spin excitations get modified from (13) - (15) as 
\be 
\frac{dS_0^x}{dt} = -[\alpha(2A+\mathcal{H}) - j](1- (S_0^z)^2) \, , 
\ee 
\be 
\frac{dS_0^y}{dt} =  (2A+\mathcal{H})S_0^z +  [\alpha(2A+\mathcal{H}) - j] S_0^xS_0^y \, , 
\ee 
\be 
\frac{dS_0^z}{dt} =  -(2A+\mathcal{H})S_0^y + [\alpha(2A+\mathcal{H})-j]S_0^xS_0^z \, . 
\ee 
Now choosing the spin current as 
\be 
j=\alpha(2A+\mathcal{H}) \, , 
\ee 
one can check that 
\be 
\frac{dS_0^x}{dt} = 0 \, , 
\ee 
\be 
\frac{dS_0^y}{dt} = (2A+\mathcal{H})S_0^z \, , 
\ee  
\be 
\frac{dS_0^z}{dt} = -(2A+\mathcal{H})S_0^z \, .  
\ee 
Consequently, the spin vector evolves as 
\be
\vec{S}_0 = \left(\sqrt{1-\hat{a}^2}, \hat{a}\cos(\Omega t+\delta), -\hat{a}\sin(\Omega t +\delta\right) , 
\ee 
where $\hat{a}=$constant and $\Omega=(2A+\mathcal{H})$, and the effect of damping is exactly offset by the 
spin current term.  Thus the spin current acts effectively as an ``external magnetic field plus 
anisotropy" and the system can generate microwave oscillations.  When $j<\alpha(2A+\mathcal{H})$, 
damping will overtake asymptotically and the spin will switch its direction. 

\section{$\mathcal{PT}$-symmetric magnetic device} 

Recently a class of synthetic magnetic nanostructures that makes use of the nature of 
loss/dissipation mechanism together with appropriate amplification (gain) process has 
been suggested by Lee, Kottos and Shapiro \cite{Kottos} to control magnetization dynamics. 
The suggested arrangement consists of two coupled nano-ferromagnetic films, $n=1,2$ 
(when separated by a spacer) in the presence of an external magnetic field along the x-axis, 
for example out-of-plane geometry (so that the z-axis is perpendicular to the films) as shown 
in Fig. 3.  
\begin{figure}[htp!]
  \begin{center}
    \includegraphics[width=0.4\linewidth]{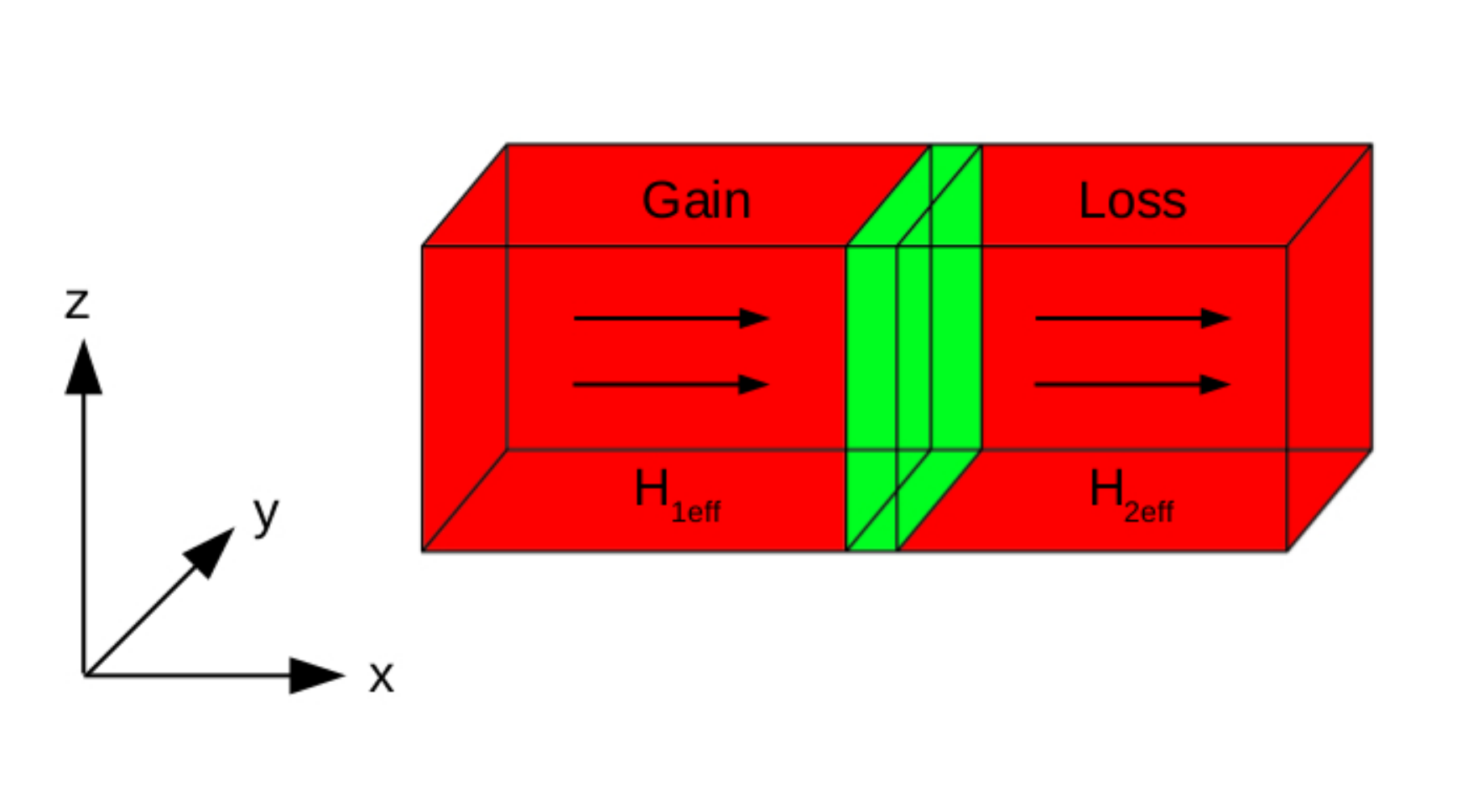}
    \caption{A $\mathcal{PT}$-symmetric trilayer structure comprising two magnetic thin films and a spacer layer suggested by Lee, Kottos and Shapiro \cite{Kottos}.}
    \label{fig:ptsymmetry}
  \end{center}
\end{figure}

\noindent Considering the effective instantaneous local fields as $\vec{\mathcal{H}}_{1eff}$ and $\vec{\mathcal{H}}_{2eff}$ 
for the two layers 1 and 2, respectively, associated with the homogeneous magnetization 
vectors $\vec{S}_1 = (\vec{M}_1/|\vec{M}_1|)$ and $\vec{S}_2=(\vec{M}_2/|\vec{M}_2|)$, 
we have the associated dynamical equations 
\be 
\frac{d\vec{S}_1}{dt} = \vec{S}_1\times\vec{\mathcal{H}}_{1eff} + k \vec{S}_1\times\vec{S}_2 + 
\alpha\vec{S}_1\times\frac{d\vec{S}_1}{dt} \, , 
\ee
\be 
\frac{d\vec{S}_2}{dt} = \vec{S}_2\times\vec{\mathcal{H}}_{2eff} + k \vec{S}_2\times\vec{S}_1 -  
\alpha\vec{S}_2\times\frac{d\vec{S}_2}{dt} \, , 
\ee 
where $k$ is the ferromagnetic coupling and $\alpha$ is the damping/gain coefficient. Note that the combined systems (41)-(42) are invariant under the simultaneous changes of the variables $\vec{S} _{1,2}\rightarrow-\vec{S} _{2,1}$, $\vec{\mathcal{H}} _{1,2\; eff}\rightarrow-\vec{\mathcal{H}} _{2,1\;eff}$ and $t \rightarrow -t$, which may be treated as equivalent to combined $\mathcal{PT}$-symmetry operation \cite{Kottos}. Now we choose the two layers such that 
\be 
\vec{S}_2 = \vec{S}_1\times\vec{S}_p \, , ~~~ \vec{S}_1 = -\vec{S}_2\times\vec{S}_p \, , 
\ee 
where $\vec{S}_p=(1,0,0)$ is a fixed polarization vector. Equation (43) implies that $\vec{S}_p$ 
is perpendicular to the plane of $\vec{S}_1$ and $\vec{S}_2$. Then, similar to the analysis in 
Sec. 4, we can choose the ferromagnetic coupling $k$ such that for simple anisotropy as 
in Eqs. (13) - (15) and external magnetic field $\mathcal{H}$, we can choose 
\be 
k = \alpha(2A+\mathcal{H}) \, , 
\ee 
so that the gain/loss terms are exactly cancelled by the ferromagnetic coupling, leaving out 
\be 
\frac{d\vec{S}_1}{dt} = \vec{S}_1\times\vec{\mathcal{H}}_{1eff} \, , 
\ee 
\be 
\frac{d\vec{S}_2}{dt} = \vec{S}_2\times\vec{\mathcal{H}}_{2eff} \, , 
\ee 
leading to spin oscillations and thereby to an effective control of magnetization oscillations.

\section{More general spin excitations} 

One can consider more general localized spin excitations like two, three, etc. spin excitations. 
For example, in the case of localized two-spin excitations, 
\bea 
\vec{S}_n &=& ..., (1,0,0), (1,0,0), (S_i^x, S_i^y,S_i^z), (S_{i+1}^x, S_{i+1}^y,S_{i+1}^z), (1,0,0), (1,0,0), ... \nonumber \\
&=& ..., (1,0,0), (1,0,0), (S_0^x, S_0^y, S_0^z), (S_1^x, S_1^y, S_1^z), (1,0,0), (1,0,0), ... \, , 
\eea 
we obtain the dynamical equations from (5) - (7) as 
\bea 
\frac{d{S}_0^x}{dt} &=& CS_0^yS_1^z - BS_0^zS_1^y - 2DS_0^yS_0^z \nonumber\\
&&+ \alpha[BS_0^{x}S_0^{y}S_1^{y}-(A(S_1^x+1)+\mathcal{H})((S_0^y)^2+(S_0^z)^2)+CS_0^{x}S_0^{z}S_1^{z}-2DS_0^{x}(S_0^{z})^{2}] \, , \\
\frac{d{S}_0^y}{dt} &=& AS_0^z(S_1^x+1) - CS_0^xS_1^z + 2DS_0^xS_0^z+\mathcal{H}S_0^z\nonumber\\
&&+ \alpha[(A(S_1^x+1)+\mathcal{H})S_0^xS_0^y-BS_1^y((S_0^x)^2+(S_0^z)^2)+CS_0^yS_0^zS_1^z-2DS_0^y(S_0^z)^2] \, , \\
\frac{d{S}_0^z}{dt} &=& BS_0^xS_1^y - AS_0^y(S_1^x+1) - \mathcal{H}S_0^y \nonumber\\
&&+ \alpha[(A(S_1^x+1)+\mathcal{H})S_0^xS_0^z+BS_0^{y}S_0^{z}S_1^{y}-CS_1^z((S_0^x)^2+(S_0^y)^2)+2DS_0^z((S_0^x)^2+(S_0^y)^2)] \, , \\
\frac{d{S}_1^x}{dt} &=& CS_0^zS_1^y - BS_0^yS_1^z - 2DS_1^yS_1^z \nonumber\\
&&+ \alpha[BS_0^{y}S_1^{x}S_1^{y}-(A(S_0^x+1)+\mathcal{H})((S_1^y)^2+(S_1^z)^2)+CS_0^{z}S_1^{x}S_1^{z}-2DS_1^{x}(S_1^{z})^{2}] \, , \\
\frac{d{S}_1^y}{dt} &=& AS_1^z(S_0^x+1) -CS_1^xS_0^z + 2DS_1^xS_1^z + \mathcal{H}S_1^z \nonumber\\
&&+ \alpha[(A(S_0^x+1)+\mathcal{H})S_1^xS_1^y-BS_0^y((S_1^x)^2+(S_1^z)^2)+CS_0^zS_1^yS_1^z-2DS_1^y(S_1^z)^2] \, , \\
\frac{d{S}_1^z}{dt} &=& BS_1^xS_0^y - AS_1^y(S_0^x+1) - \mathcal{H}S_1^y \nonumber\\
&&+ \alpha[(A(S_0^x+1)+\mathcal{H})S_1^xS_1^z+BS_0^{y}S_1^{y}S_1^{z}-CS_0^z((S_1^x)^2+(S_1^y)^2)+2DS_1^z((S_1^x)^2+(S_1^y)^2)] \, .  
\eea
Note that the terms proportional to $\alpha$ are generalizations for the present two-spin 
excitation case compared to those given in Eqs. (9) - (11).  As such the system (48) - (53) 
does not seem to be analytically solvable.  In Fig. 4, we numerically integrate the system 
for both the undamped case ($\alpha=0$) and the damped case ($\alpha \ne 0$) for 
nonzero $D$ and present the solutions in the undamped and damped cases to show the 
existence of more general internal localized excitations. The analysis can be extended 
to even more general situations, which will be presented elsewhere.  
\begin{figure}[htp!]
\centering
\subfigure{\includegraphics[width=0.32\linewidth]{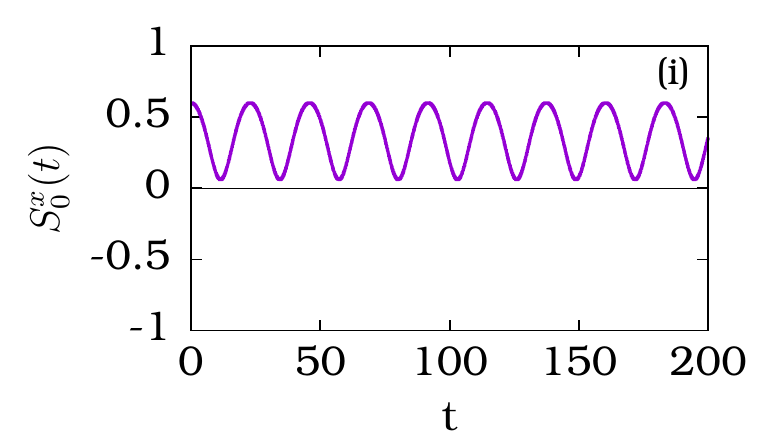}}
\subfigure{\includegraphics[width=0.32\linewidth]{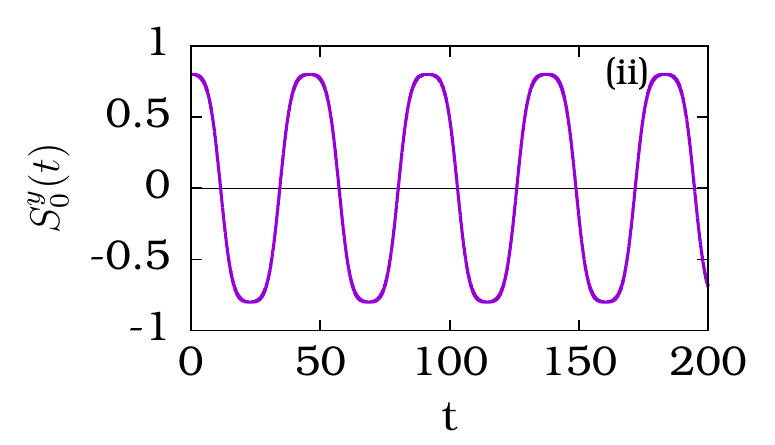}}
\subfigure{\includegraphics[width=0.32\linewidth]{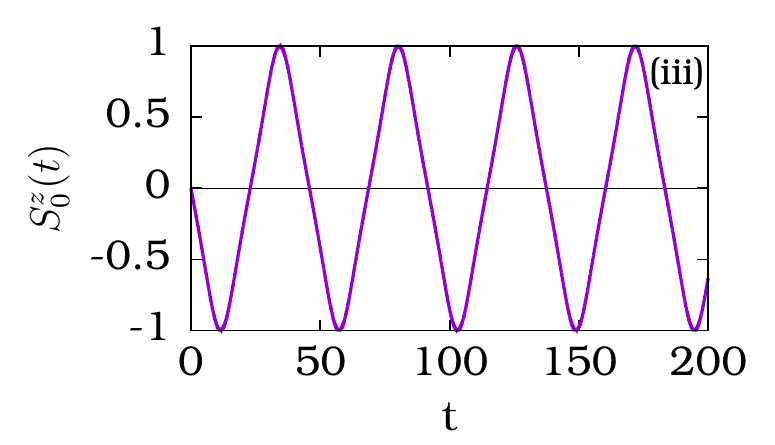}}
\subfigure{\includegraphics[width=0.32\linewidth]{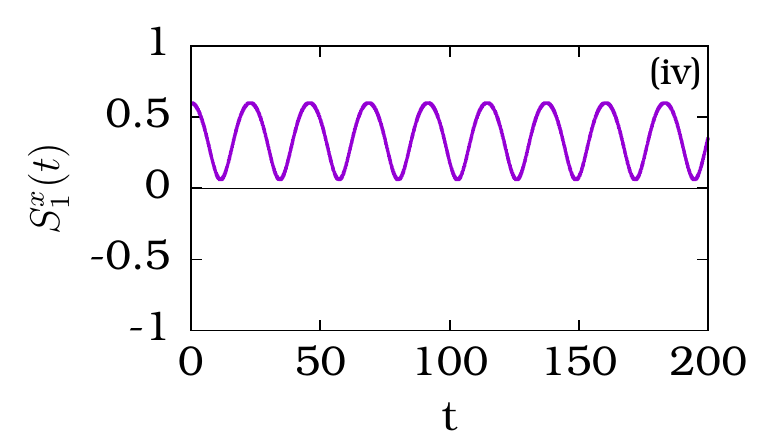}}
\subfigure{\includegraphics[width=0.32\linewidth]{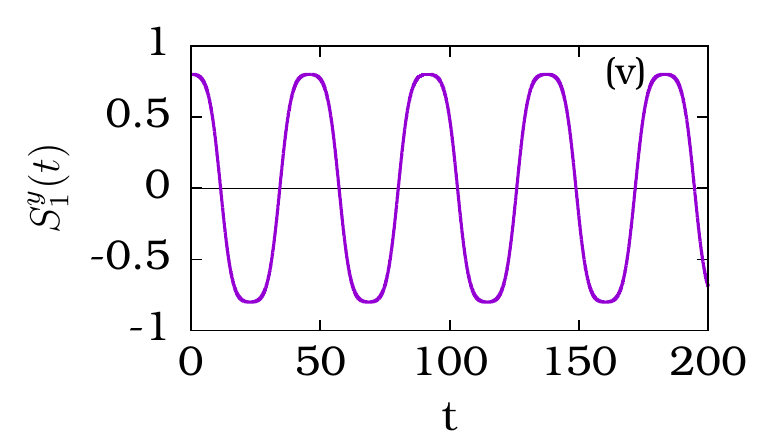}}
\subfigure{\includegraphics[width=0.32\linewidth]{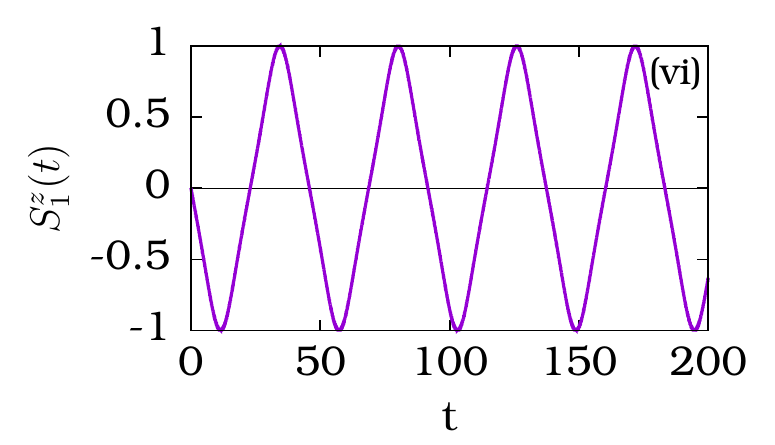}}
Fig. 4 (a): Undamped two-spin excitations
\label{fig:twospin_undamped}
\end{figure}
\begin{figure}[htp!]
\centering
\subfigure{\includegraphics[width=0.32\linewidth]{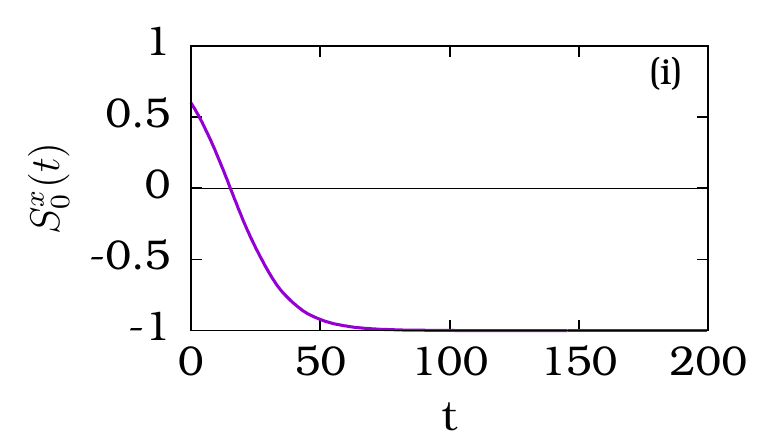}}
\subfigure{\includegraphics[width=0.32\linewidth]{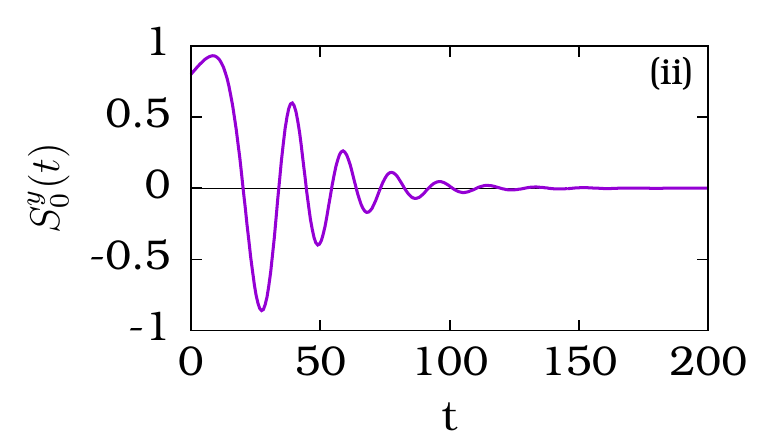}}
\subfigure{\includegraphics[width=0.32\linewidth]{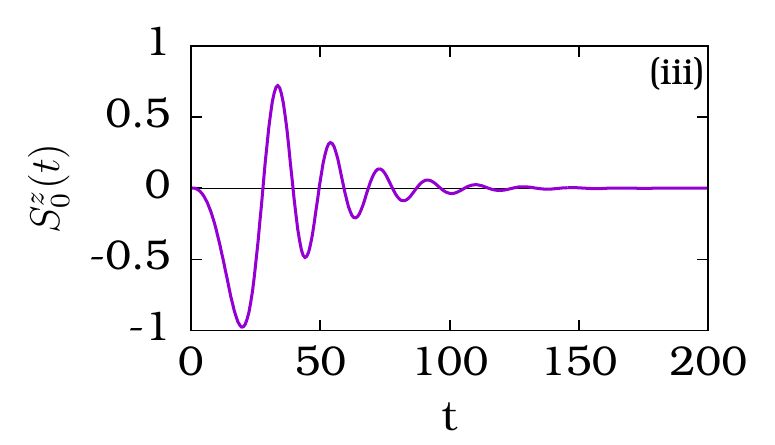}}
\subfigure{\includegraphics[width=0.32\linewidth]{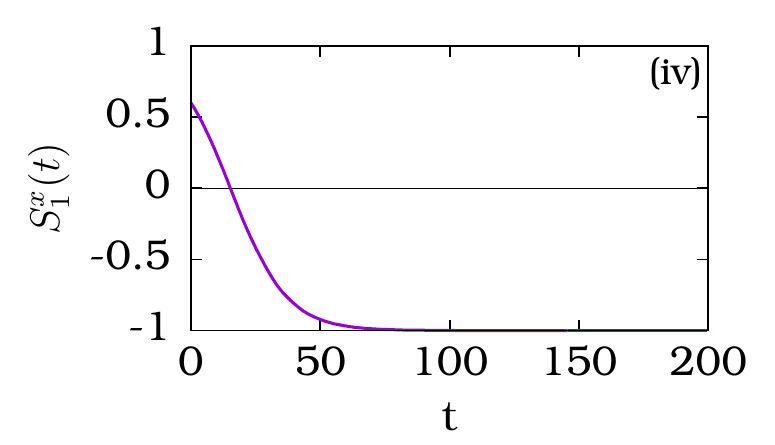}}
\subfigure{\includegraphics[width=0.32\linewidth]{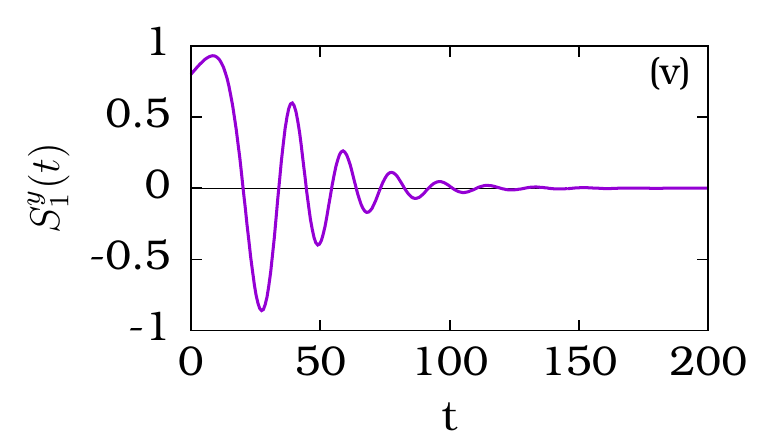}}
\subfigure{\includegraphics[width=0.32\linewidth]{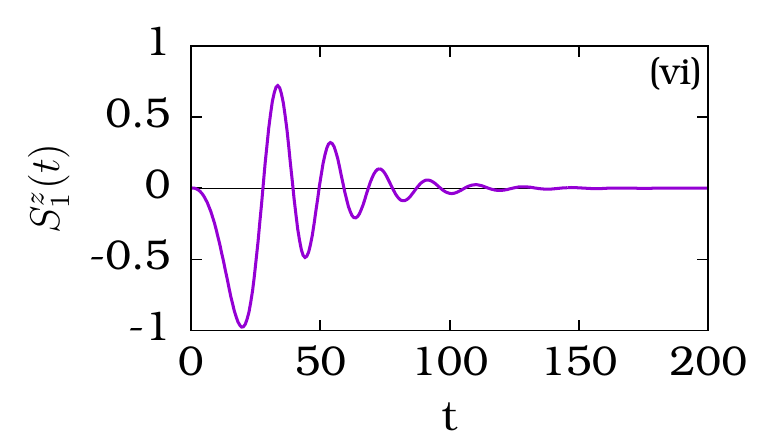}}
Fig. 4 (b): Damped two-spin excitations
\caption{Solution of Eqs. (48)-(53) for two-spin excitations $\textbf{S}_0$ and $\textbf{S}_1$ for the (a) undamped ($\alpha=0$) and (b) damped cases ($\alpha=0.005$), with the anisotropy parameters $A=0.1$, $B=0.23$, $C=1.0$ and $D=0.3$ and the magnetic field $\mathcal{H}=113\, Oe$. }
\label{fig:twospin_damped}
\end{figure}
\section{Conclusion} 

By looking at the simplest internal localized excitations in an anisotropic Heisenberg 
ferromagnetic spin chain in external magnetic field with additional Gilbert damping, we 
deduced the explicit solutions which characteristically show the effect of damping.  Then 
applying a spin current in an STNO of appropriate magnitude, we pointed out how the 
tendency toward damping can be offset exactly and thereby sustaining the magnetic 
oscillations.  Our prediction about a $\mathcal{PT}$-symmetric STNO could be tested in 
magnetic multilayer structures with carefully balanced gain and loss.  We have also pointed 
out how such controlled oscillations can be effected in a recently suggested nanomagnetic 
trilayer device. It will be insightful to observe these oscillations in appropriate magnetic 
systems experimentally. Finally, in a related context we note that nonreciprocal optical 
modes can exist at an interface between two $\mathcal{PT}$-symmetric magnetic domains 
near a frequency corresponding to almost zero effective permeability \cite{Wang}.

\section{Acknowledgments} 
The authors wish to thank Dr. D. Aravinthan for his help in the numerical analysis. 
The research work of ML was supported by a NASI Senior Scientist Platinum Jubilee 
Fellowship (NAS 69/5/2016-17) and a DST-SERB Distinguished Fellowship (No.: SERB/F/6717/2017-18). ML was also supported by a Council of Scientific 
and Industrial Research, India research project (No.: 03/1331/15/EMR-II) and a National Board for Higher Mathematics research project (No.: 2/48(5)/2015/NBHM(R.P.)/R\&D II/14127).  ML also wishes to thank the Center for Nonlinear Studies, Los Alamos National Laboratory, 
USA for its warm hospitality during his visit in the summer of 2017.  This work was 
supported in part by the U.S. Department of Energy. 

\section*{Appendix A}
Here we briefly point out how to solve Eq. (20). Introducing the transformation 
\be 
S_0^y(t) = e^{\alpha(2A+\mathcal{H})  \int S_0^x dt} \cdot \hat{S}_0^y(t) \, \tag{A. 1}
\ee 
into Eq. (20), we obtain 
\be
\frac{d^2{\hat{S}}_0^y(t)} {dt^2} + (2A+\mathcal{H})^2{\hat{S}}_0^y(t) = 0 \, .  \tag{A. 2}
\ee 
Consequently, we have 
\be 
\hat{S}_0^y(t) = \hat{a} \cos(\Omega t+\delta) \, , ~~~ \Omega = 2A+\mathcal{H} \, , \tag{A. 3}
\ee
where $\hat{a}$ and $\delta$ are arbitrary constants. Then, the prefactor on the right hand side of 
(21) can be deduced as follows.  Since 
\be 
I = \int S_0^x dt = \int \frac{c^2 \exp({-2\alpha(2A+\mathcal{H})t}) -1}{c^2 \exp({-2\alpha(2A+\mathcal{H})t})+1} dt 
=-\frac{1}{2\alpha(2A+\mathcal{H})} \log \frac{(c^2 \exp({-2\alpha(2A+\mathcal{H})t})+1)^2}{c^2 \exp({-2\alpha(2A+\mathcal{H})t}} \, , \tag{A. 4}
\ee 
the prefactor becomes 
\be 
\exp\left[\alpha(2A+\mathcal{H})\int S_0^x dt  \right] = \frac{c\exp(-\alpha(2A+\mathcal{H})t)} {c^2 \exp({-2\alpha(2A+\mathcal{H})t})+1}  \, . \tag{A. 5}
\ee 
Correspondingly 
\be 
S_0^y = \frac{c\exp(-\alpha(2A+\mathcal{H})t)} {c^2 \exp({-2\alpha(2A+\mathcal{H})t})+1}~ \hat{a}\cos(\Omega t + \delta) \,  , 
~~~ \Omega = 2A+\mathcal{H} \, ,\tag{A. 6}
\ee
which is Eq. (21).


\begin{thebibliography}{99}
\bibitem{Hiller} B. Hillerbrands and K. Ounadjela, {\it Spin Dynamics in Confined 
Magnetic Structures}, Vols. I \& II (Springer, Berlin) 2002. 
\bibitem{ML1} M. Lakshmanan, Philos. Trans. R. Soc. A {\bf 369} (2011) 1280. 
\bibitem{Fert} B. Georges, V. Cros and A. Fert, Phys. Rev. B {\bf 73} (2006) 0604R. 
\bibitem{Yang} Z. Yang, S. Zhang and Y. C. Li, Phys. Rev. Lett. {\bf 99} (2007) 134101. 
\bibitem{ML2} M. Lakshmanan, Phys. Lett. A {\bf 61} (1977) 53. 
\bibitem{Naka} K. Nakamura and T. Sasada, J. Phys. C {\bf 15} (1982) L915. 
\bibitem{Sklyanin} E. K. Sklyanin, LOMI preprint E-3-79, Leningrad (1979). 
\bibitem{Ishi} Y. Ishimori, Prog. Theor. Phys. {\bf 72} (1984) 33. 
\bibitem{ML3} M. Lakshmanan and A. Saxena, Physica D {\bf 237} (2008) 885. 
\bibitem{Zabel} H. Zabel and M. Farle (Eds.), {\it Magnetic Nanostructures: Spin Dynamics and Spin Transport} (Springer, Berlin) 2013. 
\bibitem{Sievers} A. Sievers and S. Takeno, Phys. Rev. Lett. {\bf 61} (1988) 970. 
\bibitem{Zolo} Y. Zolotaryuk, S. Flach and V. Fleurov, Phys. Rev. B {\bf 63} (2003) 214422. 
\bibitem{Subash1} M. Lakshmanan, B. Subash and A. Saxena, Phys. Lett. A {\bf 378} (2014) 1119. 
\bibitem{Bert} G. Bertotti, I. Mayergoyz and C. Serpico, {\it Nonlinear Magnetization Dynamics in 
Nanosystems} (Elsevier, Amsterdam) 2009. 
\bibitem{Georges} B. Georges, J. Grollier, V. Cros and A. Fert, Appl. Phys. Lett. {\bf 92} (2008) 232504. 
\bibitem{Subash2} B. Subash, V. K. Chandrasekar and M. Lakshmanan, Europhys. Lett. {\bf 102} 
(2013) 17010; {\bf 109} (2015) 17009. 
\bibitem{Turtle} J. Turtle, K. Beauvais, R. Shaffer, A. Palacios, V. In, T. Emery and P. Langhini, 
J. Appl. Phys. {\bf 113} (2013) 114901. 
\bibitem{LLGS} J. C. Slonczewski, J. Magn. \& Magn. Mater. {\bf 159} (1996) L261. 
\bibitem{Kottos} J. M. Lee, T. Kottos and B. Shapiro, Phys. Rev. B {\bf 91} (2015) 094416. 
\bibitem{Wang} J. Wang, H. Y. Dong, C. W. Ling, C. T. Chan and K. H. Fung, Phys. Rev. B {\bf 91} (2015) 235410. 

\end{thebibliography}
 \end{document}